\pdfoutput=1
\documentclass[11pt]{article}

\usepackage[final]{acl}
\usepackage{times}
\usepackage{latexsym}
\usepackage[T1]{fontenc}
\usepackage[utf8]{inputenc}
\usepackage{microtype}
\usepackage{inconsolata}
\usepackage{amsmath,amsthm,amssymb,graphicx}
\usepackage{multirow} 
\usepackage{booktabs}
\usepackage{pgfplots}
\usepackage{subfigure}
\usepackage{pifont}
\usepackage{url}

\title{VHASR: A Multimodal Speech Recognition System With Vision Hotwords}

\author{Jiliang Hu\textsuperscript{1}, Zuchao Li\textsuperscript{1,2}\thanks{Corresponding author. This work was supported by the National Natural Science Foundation of China (No. 62306216, No. 72074171, No. 72374161), the Natural Science Foundation of Hubei Province of China (No. 2023AFB816), the Fundamental Research Funds for the Central Universities (No. 2042023kf0133).}, Ping Wang\textsuperscript{3}, Haojun Ai\textsuperscript{1}, Lefei Zhang\textsuperscript{2}, Hai Zhao\textsuperscript{4} \\
\textsuperscript{1}Key Laboratory of Aerospace Information Security and Trusted Computing, Ministry of \\
Education, School of Cyber Science and Engineering, Wuhan University, Wuhan, China, \\
\textsuperscript{2}School of Computer Science, Wuhan University, Wuhan, China, \\
\textsuperscript{3}School of Information Management, Wuhan University, Wuhan, China, \\
\textsuperscript{4}Department of Computer Science and Engineering, Shanghai Jiao Tong University.}

\begin{document}
\maketitle

\begin{abstract}
The image-based multimodal automatic speech recognition (ASR) model enhances speech recognition performance by incorporating audio-related image. However, some works suggest that introducing image information to model does not help improving ASR performance. In this paper, we propose a novel approach effectively utilizing audio-related image information and set up VHASR, a multimodal speech recognition system that uses vision as hotwords to strengthen the model's speech recognition capability. Our system utilizes a dual-stream architecture, which firstly transcribes the text on the two streams separately, and then combines the outputs. We evaluate the proposed model on four datasets: Flickr8k, ADE20k, COCO, and OpenImages. The experimental results show that VHASR can effectively utilize key information in images to enhance the model's speech recognition ability. Its performance not only surpasses unimodal ASR, but also achieves SOTA among existing image-based multimodal ASR.\footnote{Our code is available at \url{https://github.com/193746/VHASR}}

\end{abstract}

\section{Introduction}
\begin{figure}[htbp]
	\centering 
	\includegraphics[scale=0.4]{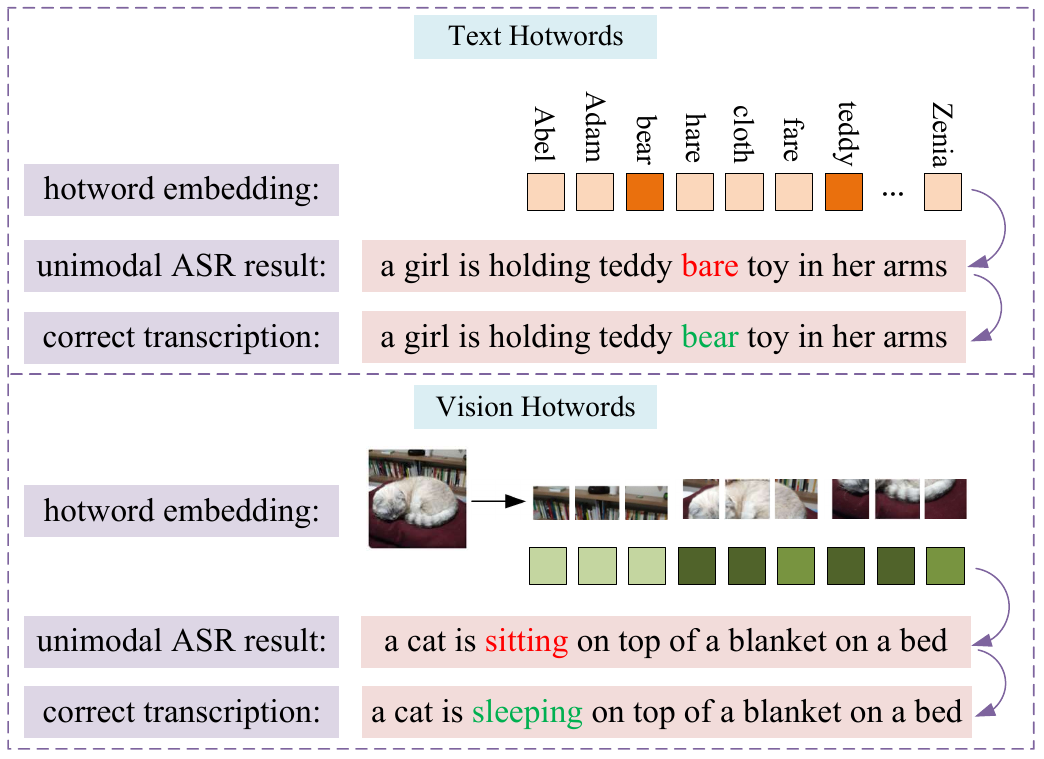} 
	\caption{Comparison between text hotwords and the vision hotwords proposed in this paper. Text hotwords are a set of custom keywords that are prone to errors, while image hotwords refer to patches of an image. The hotword with a darker rectangle indicates that it is more relevant to transcription.} 
 % and effectly utilizing the information of this hotword can help ASR better recognize its corresponding keyword. For hotwords with light rectangular, it is important to avoid their information affecting the model.
    \label{htwd}
\end{figure}

ASR model \citep{chan2015listen} takes audio as input and produces corresponding transcription. One effective method to improve the model's ASR performance is to increase both the volume of training data and the number of model parameters. We are now in the era of large language models (LLMs) \citep{brown2020language,li2023batgpt}, which have been developed across various domains \citep{yang2024batgpt,zhang2023arcgpt}. In the speech domain, there are also many LLMs that demonstrate impressive ASR capabilities \citep{chu2023qwen,radford2023robust}. However, this approach can be expensive. A more cost-effective alternative is to introduce additional information related to speech into the model. This information can be presented in either textual or visual forms. The ASR system that utilizes audio-related information from various modalities is referred to as multimodal ASR.

Hotwords, which are terms in certain professional fields or words that are easily confused with other homonyms, are common textual cues. There have been many studies on how to freely customize hotwords and improve the recall of hotwords \citep{han2021cif, shi2024seaco}. It is also possible to use captions as textual information \citep{moriya2018lstm, han2023vilas}.

Visual cues can be in the form of video or image. Audio-Visual Speech Recognition (AVSR) enhances the accuracy of speech recognition by capturing lip movement information of characters in video \citep{ ivanko2023review}. Image-based multimodal ASR extracts visual feature from image associated with speech to correct transcription errors. We abbreviate image-based multimodal ASR as IBSR. Because the lip movement information of video's role is closely linked to his speech, it influences nearly every word in the transcribed text. In contrast, IBSR only impacts a subset of the words as the image is only associated with specific audio clips \citep{ oneațua2022improving}. IBSR currently lacks a universal and effective method for utilizing image information, leading to various experimental results in different studies. Some works \citep{sun2016look, srinivasan2020looking, srinivasan2020multimodal} have a positive effect by incorporating image information, while others \citep{srinivasan2020fine, oneațua2022improving, han2023vilas}, have the opposite effect. 

% Compared to AVSR, IBSR is less noticeable. In fact, images related to audio can help the model capture keywords in the audio, thereby correcting transcription errors in ASR.	

In this paper, we propose a novel approach effectively utilizing audio-related image information and set up VHASR, a multimodal speech recognition system that utilizes vision hotwords to enhance the model's speech recognition capability. It calculates the similarity between different modalities to improve the effectiveness of cross-modal fusion. Drawing inspiration from text hotwords, we utilize Vision Transformer (ViT) to partition images into multiple visual tokens and consider each visual token as a vision hotword. Our system adopts a dual-stream architecture. One stream is the ASR stream, which receives audio information and produces transcribed text. The other stream is the vision hotwords (VH) stream, which receives vision hotwords and audio hidden features, and generates corresponding text. In the VH stream, we calculate the similarity between audio and vision hotwords to reduce the weight of vision hotwords with low similarity. This process helps to extract fine-grained image information. When inferring, VHASR first transcribes the text separately from the ASR stream and the VH stream, and then merges the outputs. We ensure the high accuracy of the merged output by comparing the similarity of different modalities. Specifically, we first calculate the audio-image similarity to discard the VH stream if the similarity is low. Then, we calculate the image-text token similarity to compare the ASR stream and VH stream outputs by tokens. Finally, tokens with higher similarity are selected for the merged output. 

We evaluate the proposed model on four datasets: Flickr8k, ADE20k, COCO, and OpenImages. The experimental results show that VHASR can effectively utilize critical information in images to improve the model's ASR performance. Its performance is not only better than ordinary unimodal ASR models but also surpasses existing IBSR models. The contributions of this paper are as follows:

\begin{enumerate}
	\item[(1)] We demonstrate that through our idea of vision hotwords, injecting audio-related image into the ASR model can help the model correct transcription errors.
	\item[(2)] We propose VHASR, by utilizing a dual-stream architecture and calculating the cross-modal similarity, it promotes effective utilization of visual information in vision hotwords.
	\item[(3)] The proposed model achieves SOTA on Flickr8k, ADE20k, COCO, and OpenImages.

 % calculating the similarity between different modalities for reduction and classification, it can promote effective fusion of cross-modal information
\end{enumerate}

\section{Related Work}
\textbf{Image-based multimodal ASR.} \citet{sun2016look} introduce a multimodal speech recognition scenario which utilizes images to assist the language model in decoding the most probable words and rescoring the top hypotheses. \citet{ caglayan2019multimodal} propose an end-to-end multimodal  ASR system implemented by LSTM \citep{graves2012long}. They apply visual adaptive training \citep{palaskar2018end} to finetune a pretrained ASR model with visual data, and leverage visual information to initialize model's encoder and decoder. \citet{ srinivasan2020fine} present a model for multimodal ASR that utilizes visual feature from object proposals. They integrate the features of object proposals into a visual representation by utilizing their attention distribution as weights, and incorporate this visual representation into the model via a hierarchical attention mechanism. \citet{oneațua2022improving} combine speech and visual embeddings using two fusion approaches. One approach fuses along the embedding dimension, and another fuses along the sequence dimension. They find that the first method performs better. \citet{han2023vilas} propose a novel multimodal ASR model called ViLaS, which is based on the continuous integrate-and-fire (CIF) mechanism \citep{dong2020cif}. It can integrate image and caption information simultaneously or separately to facilitate speech recognition. \citet{chang2023multimodal} propose a multimodal ASR system for embodied agents. Their model is based on Transformer \citep{vaswani2017attention}, where the visual feature vector is concatenated to the decoder's input word embedding at every timestep of generation.

\begin{figure*}[htbp]
	\centering 
	\includegraphics[scale=0.7]{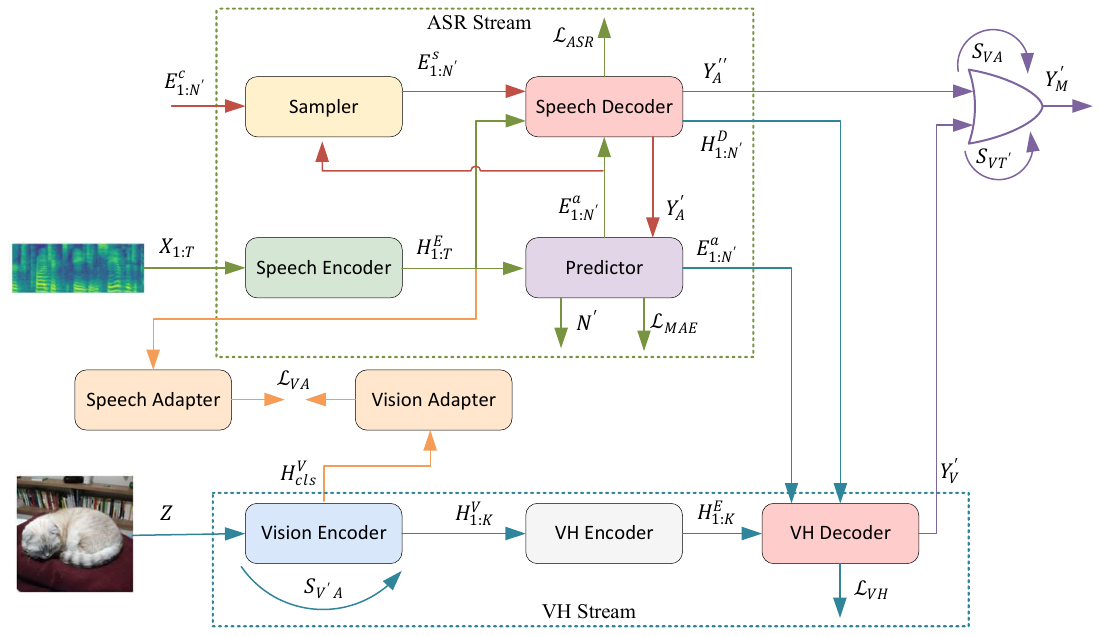} 
	\caption{The structure of our proposed model, VHASR. The green dashed box contains the modules of the ASR stream, while the blue dashed box contains the modules of the VH stream. The data flow in the ASR part is indicated by green and red lines. It only passes through the red lines during ASR model's second pass of training. The VH stream's data flow is denoted by blue lines. The data flow for calculating audio-image similarity is represented by yellow lines. The purple lines illustrate the data flow when merging two streams.} 
    \label{framework}
\end{figure*}

\noindent \textbf{Function of image information.} \citet{ srinivasan2020looking} conduct the experiment called audio corruption, in which they mask the words related to nouns and places with silence and white noise, respectively. The study demonstrates that visual representations help in recovering words that are masked in the input acoustic signal. \citet{srinivasan2020multimodal} think the previous work has only masked a fixed set of words in the audio, which is an unrealistic setting. So, they propose a method called RandWordMask, where masking can occur for any word segment to improve the audio corruption experiment. \citet{kumar2023visual} propose two effective ASR error correction methods: one employs a gated fusion method to concatenate visual and textual features, while the other utilizes image's caption as correction model's prompt. Both methods demonstrate that visual information helps restoring incorrect words in transcription. In short, image information helps to recover incorrect words in transcription that are caused by masked acoustic signals or ASR model's error.

\section{VHASR}
\subsection{ASR Stream}
Follow \citet{gao2022paraformer}, we adopt this parallel Transformer for non-autoregressive end-to-end speech recognition as the basic framework of our ASR stream. As shown in green dashed box of Figure \ref{framework}, the adopted framework consists of four parts: speech encoder, predictor, sampler, and decoder. The framework adopts two-pass training and one-pass inference. 
\subsubsection{Acoustic Representation Learning}

Let $X$ be a speech sequence with $T$ frames, $X=\{x_1,x_2,x_3,…,x_T\}$. $Y$ is a sequence of tokens, and its length is $N$. Each token is in the vocabulary $V$, $Y=\{y_1,y_2,y_3,…,y_N \mid y_i\in V\}$. 

The speech encoder adopts the SAN-M \citep{gao2020san} structure, which is a special Transformer Layer that combines self-attention mechanism with deep feed-forward sequential memory networks (DFSMN). It converts the input $X_{1:T}$ to the  hidden representation $H_{1:T}^E$.
$$
H_{1:T}^E=\textrm{SpeechEncoder}(X_{1:T})
$$

The predictor is a two-layer Deep Neural Networks (DNN) model that aligns speech and text based on CIF. It is used to predict the length of sentences $N^{'}$ and extract acoustic representation $E_{1:N^{'}}^a$ from the speech encoder's hidden representation $H_{1:T}^E$. 

$$
N^{'},E_{1:N^{'}}^a=\textrm{Predictor}(H_{1:T}^E)
$$

The sampler does not contain learnable parameters and is only applied when training. It strengthens acoustic representation to semantic representation by incorporating text features, aiming to better train the context modeling ability of the speech decoder. The $E_{1:N^{'}}^c$ denotes the embedding of $Y$. The sampler initially identifies tokens in $Y_{A}^{'}$ with transcription errors, and subsequently combines the correct embeddings of these error tokens in $E_{1:N^{'}}^c$ into $E_{1:N^{'}}^a$ to generate the semantic features $E_{1:N^{'}}^s$. Not every error token's correct embedding will be incorporated into $E_{1:N^{'}}^a$, this is determined by the mixing ratio $\lambda$, $\lambda \in(0,1)$. 

% the sampler selects a specific number of text vectors $E_{1:N^{'}}^c$ depending on the distance between the initial decoding result $Y_{A}^{'}$ and the reference text $Y$. It then combines the acoustic feature  $E_{1:N^{'}}^a$ with the text feature $E_{1:N^{'}}^c$ to create semantic feature $E_{1:N^{'}}^s$ that incorporate semantic information.

% $$
% E_{1:N^{'}}^s=\textrm{Sampler}(Y_{A}^{'},E_{1:N^{'}}^a,E_{1:N^{'}}^c)
% $$

$$
E_{1:N^{'}}^s=\textrm{Sampler}(E_{1:N^{'}}^a,E_{1:N^{'}}^c,\lceil \lambda \sum_{i=1}^{N^{'}}(y_{i}^{'} \neq y_{i}) \rceil)
$$

\subsubsection{Decoding Process}
The speech decoder adopts the bidirectional SAN-M structure. In the first pass of training, the hidden representation $H_{1:T}^E$ obtained by the speech encoder and the acoustic representation $E_{1:N^{'}}^a$ generated by the predictor are input to the speech decoder to obtain the initial decoding result $Y_{A}^{'}$.

$$
Y_{A}^{'}=\textrm{SpeechDecoder}(H_{1:T}^E,N^{'},E_{1:N^{'}}^a)
$$

In the second pass of training, the hidden representation $H_{1:T}^E$ and the semantic representation $E_{1:N^{'}}^s$ obtained by the sampler are input to the speech decoder to obtain the second decoding result $Y_{A}^{''}$
$$
Y_{A}^{''}=\textrm{SpeechDecoder}(H_{1:T}^E,N^{'},E_{1:N^{'}}^s)
$$

During the first pass, no gradient backpropagation is performed, and $Y_{A}^{'}$ is only used to determine the sampling number of the sampler. $Y_{A}^{''}$ obtained in the second pass is used to calculate the ASR loss. In inference, the model directly takes $Y_{A}^{'}$ as output and does not calculate $Y_{A}^{''}$.

\subsection{Vision Hotwords Stream}
\subsubsection{Vision Representation Learning}
\label{fine-grained}

In the VH stream, we need to extract visual features from images by the vision encoder firstly. A naive idea is to extract the features from the entire image. Because most of the information in the image is unrelated to the audio, especially the background of the image. The introduction of irrelevant information may cause the visual features to become noise. Therefore, we should consider a strategy to extract fine-grained image information.

The vision encoder is essentially ViT \citep{dosovitskiy2020image}. ViT uses Transformer to extract visual features. It follows the application of the Transformer in natural language processing by initially dividing the image into multiple patches, considering each patch as a token, embedding the positional information, and then feeding visual tokens \citep{peng2024multi} into the Transformer. The features outputted by ViT are the features of each visual token. If the downstream task of ViT is classification, a trainable {\tt CLS} token can be added in front of the visual token. The score on the {\tt CLS} token can then be utilized for classification. It would be a good choice if we utilize each visual tokens' features instead of entire image's features. At the token granularity level, we can diminish the impact of tokens unrelated to audio and amplify the influence of tokens related to audio.

So, our strategy is to calculate the features of each visual token and then adjust the weight of visual tokens. For the ASR model with text hotwords, it is often necessary to consider how to capture involved hotwords and exclude unrelated hotwords when there are many customized hotwords. This is similar to our consideration, so we call each visual token an vision hotword. Let Z be the input image. First, utilize the vision encoder to transform it into token-level visual features $H_{0:K}^{V}$, where K represents the number of vision hotwords. The initial features of $H_{0:K}^{V}$, corresponds to the features of the {\tt CLS} token, while others are vision hotwords' features.

$$
H_{0:K}^{V}=\textrm{VisionEncoder}(Z)
$$
$$
H_{CLS}^{V}=H_{0:K}^{V} \lbrack \, 0 \, \rbrack \: ; H_{1:K}^{V}=H_{0:K}^{V} \lbrack \, 1 \!: \! K \, \rbrack
$$
% $$
% H_{0:K}^{V}=\{ H_{CLS}^{V},H_{1:K}^{V} \}
% $$

We determine the correlation between each vision hotword and audio by calculating their cosine similarity. Specifically, the first step is to input $H_{1:K}^{V}$ into the vision adapter, which is composed of a linear layer, to obtain $H_{1:K}^{V^{'}}$. Next, we add the embedding of a trainable {\tt CLS} token to the beginning of the acoustic features $H_{1:T}^{E}$, resulting in $H_{0:T}^{E}$. This $H_{0:T}^{E}$ is then fed into the speech adapter, which consists of a Transformer layer, to produce the complete audio features $H^{E^{'}}$.

$$
H_{1:K}^{V^{'}}=\textrm{VisionAdapter}(H_{1:K}^{V})
$$
$$
H^{E^{'}}=\textrm{SpeechAdapter}(H_{0:T}^{E})[0]
$$

Then, calculate cosine similarity between vision hotwords and audio, denoted as $S_{V^{'}A}$.

$$
S_{V^{'}A}=\textrm{cos}(H_{1:K}^{V^{'}},H^{E^{'}})
$$

Finally, we adjust the weight of $H_{1:K}^{V}$ by $S_{V^{'}A}$.

$$
H_{1:K}^{V}=H_{1:K}^{V} \times S_{V^{'}A}
$$

\begin{figure}[htbp]
	\centering 
	\includegraphics[scale=0.5]{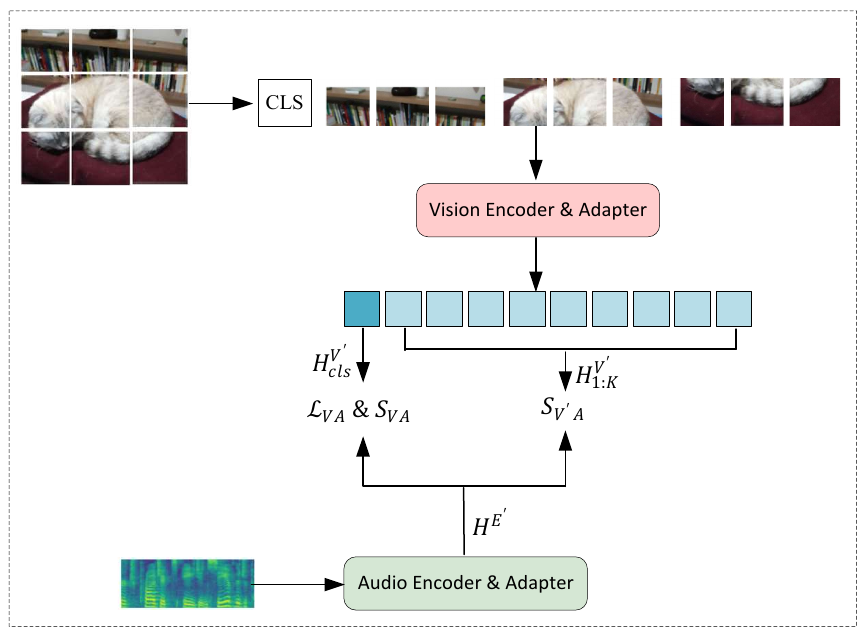} 
	\caption{Using vision hotword-audio similitude and image-audio similitude to learn fine visual representation.} 
    \label{sva}
\end{figure}

In order to enhance the effectiveness of similarity-based weight adjustment, an additional loss needs to be introduced to train the adapters. We utilize the acoustic features and the {\tt CLS} token's features of the image to calculate the image-audio contrastive loss $\mathcal{L}_{VA}$ to optimize the adapters. The reason for using image-audio contrastive loss instead of vision hotwords-audio contrastive loss is that the former has a coarser granularity, making it easier to converge. Moreover, during inference, we need to use image-audio similarity for decoding optimization, which will be explained at length in Section \ref{merging}. Figure \ref{sva} illustrates in detail our optimization of visual representation by calculating the similitude between vision hotwords and audio, as well as the similitude between image and audio.

$$
H_{CLS}^{V^{'}}=\textrm{VisionAdapter}(H_{CLS}^{V})
$$
$$
\mathcal{L}_{VA}=\textrm{ContrastiveLoss}(H_{CLS}^{V^{'}},H^{E^{'}})
$$

\subsubsection{Decoding Process}
The blue line in Figure \ref{framework} illustrates the data flow of the VH module. After extracting the fine visual representation of $H_{1:K}^{V}$, we further refine it using an LSTM-based VH encoder to obtain $H_{1:K}^{E}$.

$$
H_{1:K}^{E}=\textrm{VHEncoder}(H_{1:K}^{V})
$$

The next step is to use a text decoder to obtain the probability distribution of each token. Obviously, if we only use $H_{1:K}^{E}$ which just contains image information as input, it will result in a significant deviation in the probability distribution of tokens, and the VH stream's outcome will be completely inconsistent with the correct transcription. So, we need to incorporate certain hidden features of the ASR stream to modify the output of the VH stream. Drawing lessons from the idea of \citet{shi2024seaco}, we integrate the acoustic features vector $E_{1:N^{'}}^{a}$ outputted by the predictor and the hidden features $H_{1:N^{'}}^{D}$ outputted by the speech decoder with $H_{1:K}^{E}$ separately to derive $E_{1:N^{'}}^{a^{'}}$ and $H_{1:N^{'}}^{D^{'}}$, which have been influenced by image information. The VH decoder adopts the same bidirectional SAN-M architecture as the speech decoder.

$$
E_{1:N^{'}}^{a^{'}}=\textrm{VHDecoder}(E_{1:N^{'}}^{a},H_{1:K}^{E})
$$
$$
H_{1:N^{'}}^{D^{'}}=\textrm{VHDecoder}(H_{1:N^{'}}^{D},H_{1:K}^{E})
$$

The final input to the VH output layer is the average of $E_{1:N^{'}}^{a^{'}}$ and $H_{1:N^{'}}^{D^{'}}$.

$$
Y_{V}^{'}=\underset{y_i\in V}{\arg\max}(W_{1:V}^V \frac{(E_{1:N^{'}}^{a^{'}}+H_{1:N^{'}}^{D^{'}})}{2}+b_{1:V}^V)
$$

\subsection{Dual-stream Merging}
\label{merging}
\begin{figure}[htbp]
	\centering 
	\includegraphics[scale=0.6]{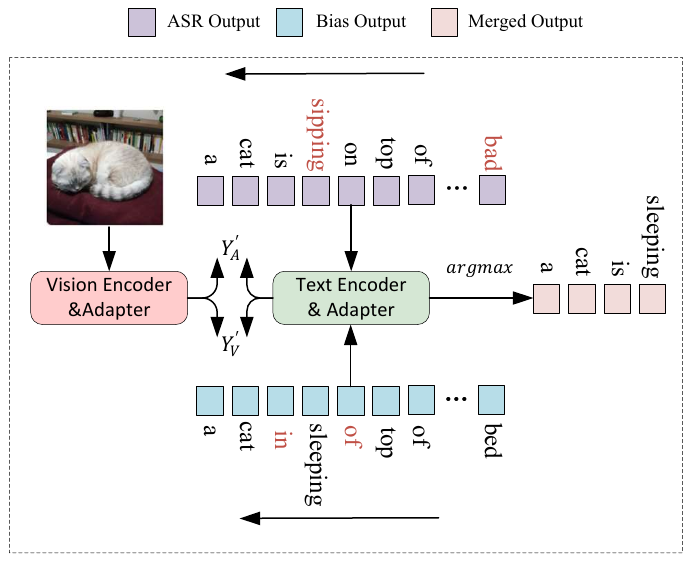} 
	\caption{The specific process of decoding optimization.} 
    \label{mge2}
\end{figure}

In this section, we will discuss how to merge the outputs of the ASR stream and the VH stream. A straightforward approach is to add the probability distributions of tokens from two modules by assigning a specific weight, denoted as $M_1$. The formula for $M_1$ is as follows, where $p_A$, $p_V$, and $p_M$ are the tokens' probability distributions of the ASR stream, VH stream, and merged result. $\alpha$ is the proportion of $p_A$, and $\alpha \in(0,1)$.
$$
p_M=\alpha \textrm{Softmax}(p_A)+(1-\alpha) \textrm{Softmax}(p_V)
$$
$$
Y_{M_{1}}^{'}=\underset{y_i\in V}{\arg\max}(p_M)
$$

The $M_1$ has low flexibility, making it difficult to achieve good results in practice. Figure \ref{mge2} illustrates a merging method based on image-token similarity, referred to as $M_2$. The vision encoder and adapter are used to calculate the visual features of the image, $H_{CLS}^{V^{'}}$, and the text encoder and adapter are used to calculate the features of each token, $H_{1:N^{'}}^{T^{'}}$. The formula for $H_{CLS}^{V^{'}}$ has been provided in Section \ref{fine-grained}, and the formula for $H_{1:N^{'}}^{T^{'}}$ is as follows. The text encoder consists of Transformer layers, the text adapter consists of a linear layer, and $Embedding$ is a additional embedding layer.
$$
H_{1:N^{'}}^{T}=\textrm{TextEncoder}(\textrm{Embedding}(Y^{'}))
$$
$$
H_{1:N^{'}}^{T^{'}}=\textrm{TextAdapter}(H_{1:N^{'}}^{T})
$$

Based on $H_{CLS}^{V^{'}}$ and $H_{1:N^{'}}^{T^{'}}$, the cosine similarity of the image and tokens, $S_{VT^{'}}$, can be calculated.
$$
S_{VT^{'}}=\textrm{cos}(H_{CLS}^{V^{'}},H_{1:N^{'}}^{T^{'}})
$$

When calculating $Y_{M_{2}}^{'}$, we first calculate the text features of the ASR stream output $Y_{A}^{'}$ and the VH stream output $Y_{V}^{'}$, respectively, namely $H_{1:N^{'}}^{T_{A}^{'}}$ and $H_{1:N^{'}}^{T_{V}^{'}}$. Then calculate their cosine similarities with $H_{CLS}^{V^{'}}$ separately, namely $S_{VT^{'}}^{A}$ and $S_{VT^{'}}^{V}$. Finally, a token by token comparison of the dual-stream is conducted according to $S_{VT^{'}}^{A}$ and $S_{VT^{'}}^{V}$. Specifically, the value of these two similarities at any position represents the similarity score between the token at that position and the image. At the same position, $Y_{A}^{'}$ and $Y_{V}^{'}$ may obtain different tokens. We determine which token to choose as the final result by judging the value of $S_{VT^{'}}^{A}$ and $S_{VT^{'}}^{V}$ at that position. If $S_{VT^{'}}^{A}>S_{VT^{'}}^{V}$, we take the token on $Y_{A}^{'}$, and vice versa. After completing $N^{'}$ comparisons, $Y_{M_{2}}^{'}$ can be obtained. 

% Note that because the lengths of $Y_{A}^{'}$ and $Y_{V}^{'}$ are both $N^'$, the lengths of $S_{VT^{'}}^{A}$ and $S_{VT^{'}}^{V}$ must also be $N^'$, and there will be no length mismatch.

In Section \ref{fine-grained}, to achieve an fine-grained visual representation, we additionally introduce speech and vision adapters in VHASR to compute the similarity between vision hotwords and audio. Then, to train the adapter, we calculate contrastive loss between the image and audio. In the inference stage, we can further utilize the trained adapter to optimize $M_2$ by calculating image-audio similarity. Specifically, we calculate the image-audio similarity $S_{VA}$ for a batch of data. If the audio of a piece of data does not match its own image, it is considered that the correlation between this image and audio is low. Therefore, for this data, the output of the VH stream is discarded, and the output of the ASR stream is directly used as the final output. We introduce a novel merging method called $M_3$. It involves initially filtering data with low image and audio correlation using $S_{VA}$, followed by dual-stream merging as outlined in $M_2$. We will conduct a detailed comparative experiment on these three merging methods in Section \ref{Experiment}.

\section{Experiment}
\label{Experiment}
\subsection{Configuration}
Table \ref{datasets} shows all the datasets used in this paper, with Flickr8k, ADE20k, COCO, and OpenImages used for training and testing, and SpokenCOCO used for pre-training. Flickr8k is from \citet{harwath2015deep} and SpokenCOCO is from \citet{hsu2021text}. ADE20k, COCO and OpenImages are from Local Narratives proposed by \cite{harwath2016unsupervised}. In order to shorten the experimental period, we filter data with audio exceeding 40s in ADE20k, and with more than 40 tokens or an audio duration of more than 20 seconds in COCO and OpenImages. We use word error rate (WER) as an evaluation metric to evaluate the speech recognition performance of ASR stream, VH stream, $M_1$, $M_2$, and $M_3$.

\begin{table}[htbp]
    \renewcommand{\arraystretch}{1.1} % 调整行距
	\small  
	% \resizebox{\columnwidth}{!}{
		\begin{tabular}{lccc}	 
			\hline
			Dataset&Train&Validation&Test \\
			\hline		
			Flickr8k&30,000&5,000&5,000 \\
			ADE20k&17,067&1,672&- \\
			COCO&49,109&3,232&-\\	
            OpenImages&269,749&27,813&-\\
            SpokenCOCO&592,187&25,035&-\\
			\hline		
		\end{tabular}
	% }
	\centering
	\caption{Datasets used in experiments.}
    \label{datasets}
\end{table}

\begin{table*}[htbp] 
    \renewcommand{\arraystretch}{1.1} % 调整行距
    \small %调整字体大小
	% \resizebox{\linewidth}{!}{
		\begin{tabular}{ccccccccc}	 
			\hline
			\multicolumn{2}{l}{\multirow{2}{*}{Dataset}} &Baseline&\multicolumn{6}{c}{VHASR} \\
			\cmidrule(lr){3-3}\cmidrule(lr){4-9}
			&& $\mathrm{WER}$ ($\downarrow$) & $\mathrm{Pretrain}$ & $\mathrm{WER_{ASR}}$ ($\downarrow$) &$\mathrm{WER_{VH}}$ ($\downarrow$) & $\mathrm{WER_{M_1}}$ ($\downarrow$) & $\mathrm{WER_{M_2}}$ ($\downarrow$)& $\mathrm{WER_{M_3}}$ ($\downarrow$) \\
			\hline
			
			\multicolumn{2}{l}{\multirow{2}{*}{Flickr8k}} &\multirow{2}{*}{3.86}&\ding{53}&3.84&3.94&3.82&3.62&3.60  \\
            &&& \ding{51}&3.55&3.51&3.54&3.22&\textbf{3.21} \\
            \hline
            
			\multicolumn{2}{l}{\multirow{2}{*}{ADE20k}} &\multirow{2}{*}{10.51}&\ding{53}&10.33&10.52&10.38&9.80&9.60  \\
            &&& \ding{51}&10.27&10.37&10.32&9.62&\textbf{9.53} \\
            \hline
            
            \multicolumn{2}{l}{\multirow{2}{*}{COCO}} &\multirow{2}{*}{10.44}&\ding{53}&10.35&10.34&10.28&9.63&9.61  \\
            &&& \ding{51}&10.25&10.36&10.28&9.60&\textbf{9.59}  \\			
            \hline

            \multicolumn{2}{l}{\multirow{2}{*}{OpenImages}} &\multirow{2}{*}{8.72}&\ding{53}&8.61&8.58&8.58&7.73&7.71  \\
            &&& \ding{51}&8.58&8.63&8.59&7.70&\textbf{7.68} \\
            \hline
		\end{tabular}
	% }
	\centering
    \caption{Main results of proposed model in four datasets. The $\mathrm{WER_{ASR}}$ and $\mathrm{WER_{VH}}$ represent the result of the ASR stream and VH stream, respectively. $\mathrm{M_1}$ combines the outcomes of two streams with designated weights, whereas $\mathrm{M_2}$ merges by assessing the similarity between image and text tokens. Building on $\mathrm{M_2}$, $\mathrm{M_3}$ evaluates the similarity between images and audio to eliminate unrelated images.}
    % In the proposed method, $\mathrm{WER_{ASR}}$ represents the result of the ASR stream, $\mathrm{WER_{VH}}$ indicates the result of the VH stream. $\mathrm{WER_{M_1}}$, $\mathrm{WER_{M_2}}$, and $\mathrm{WER_{M_3}}$ are the results of three merging methods, respectively.
    \label{main results}
\end{table*}

Our baseline is 220M English Paraformer. In Flickr8k, we compare our model with Acoustic-LM-RNN proposed by \citet{sun2016look}, model utilizing object features as visual information (abbreviated as Multimodal (object) in the paper) from \citet{srinivasan2020looking}, Weighted-DF in \citet{srinivasan2020multimodal}, MAG proposed by \citet{srinivasan2020fine}, model fusing the two modalities along the sequence dimension (abbreviated as Multimodal (emb) in the paper) from \citet{oneațua2022improving} and ViLaS in \citet{han2023vilas}.

The modules in CLIP-Base \citep{radford2021learning} is utilized to construct the vision encoder and vision adapter for the VH stream, as well as the vision encoder and text encoder for $M_2$. The vision module of the VH stream freeze parameters during training, and the $M_2$'s modules do not require training. The 220M English Paraformer is chosen as the foundational framework for ASR stream, initialized with the same parameters as the baseline. $\lambda$ of sampler is set to 0.75 and $\alpha$ of $M_1$ is set to 0.5. The experimental environment is constructed using Funasr \citep{gao2023funasr} and ModelScope. We trained the models until convergence, and consistently utilize the Adam optimizer with a learning rate of 5e-5.

\subsection{Main Result}

Table \ref{main results} presents the results of the proposed method and baseline on four datasets. For the ASR stream and VH stream, the WER of the ASR stream is lower. The VH stream can acquire the ability of transcribing by utilizing the hidden layer's features of the ASR stream as VH decoder's input. Among the three merge methods, $M_3$ has the best results, followed by $M_2$, and finally $M_1$. This is consistent with our expected results. $M_1$ has limited flexibility, and the fixed weight proportion is not applicable to all data. By calculating image-token similarity, comparing the results of the ASR stream and VH stream token by token, and resulting in a final output with the highest similarity, $M_2$ achieves WER that are better than both $\textrm{WER}_{\textrm{ASR}}$ and $\textrm{WER}_{\textrm{VH}}$. Furthermore, by calculating audio-image similarity in addition and excluding the VH stream with low similarity, $M_3$ reduces the transcription error compared to $M_2$. For the baseline and ASR stream, ASR stream performs better, indicating that joint training of the ASR stream, VH stream, and audio-image pairing improves the unimodal ASR's performance. For the baseline and $M_3$, $M_3$ outperforms the baseline on all four datasets, demonstrating the effectiveness of our method. In addition, pre-training with large-scale corpora can further strengthen the performance of the model. We use SpokenCOCO, which contains the largest amount of data, to pre-train the proposed model, resulting in improvements in all five metrics of the model across all four datasets.

% For the baseline and ASR stream, there is not much difference in WER between the two, indicating that joint training of the ASR stream, VH stream, and audio-image pairing does not significantly affect the performance of the original unimodal ASR.

% 使所有后面的页面都达到该页面上材料的自然高度;不 橡胶垂直长度将被拉伸。
% 避免左栏剩下空间不足以容纳下一个章节标题导致的段落强行拉伸，大量留白
\raggedbottom 

\subsection{Ordinary Multimodal Fusion vs Hotword Level Multimodal Fusion}

\begin{table}[htbp] 
    \renewcommand{\arraystretch}{1.5} % 调整行距
    \small %调整字体大小
	\resizebox{\linewidth}{!}{
		\begin{tabular}{cccccc}	 
			\hline
			\multicolumn{4}{l}{\multirow{2}{*}{Model}} &\multicolumn{2}{c}{Word Error Rate ($\downarrow$)} \\
			\cmidrule(lr){5-6}
			&&&& w/o vision & w vision \\
			\hline
			
			\multicolumn{4}{l}{Acoustic-LM-RNN \citep{sun2016look}} &14.75&13.81 ($\downarrow$ 0.94) \\
			\multicolumn{4}{l}{Multimodal (object) \citep{srinivasan2020looking}} &16.40&14.80 ($\downarrow$ 1.60) \\
			\multicolumn{4}{l}{Weighted-DF \citep{srinivasan2020multimodal}} &13.70&13.40 ($\downarrow$ 0.30) \\
            \multicolumn{4}{l}{MAG \citep{srinivasan2020fine}} &13.60&13.80 ($\uparrow$ 0.20)\\
			\multicolumn{4}{l}{Multimodal (emb) \citep{oneațua2022improving}} &3.80&4.30 ($\uparrow$ 0.50) \\
            \multicolumn{4}{l}{ViLaS \citep{han2023vilas}} &3.40&3.40 ($\downarrow$ 0)\\
            
			\hline
			\multicolumn{4}{l}{VHASR} &3.86&\textbf{3.21} ($\downarrow$ 0.65) \\	
			\hline
						
		\end{tabular}
	}
	\centering
    \caption{Comparison results with benchmarks in F8k.}
    % The languae model (LM) in \citet{sun2016look} is image-LM when using vision information. Multimodal (object) in \citet{srinivasan2020looking} refers to utilizing object features as vision information. Weighted-DF in \citet{srinivasan2020multimodal} fuses the acoustic features and vision features with a timestep dependent weighted scalar. MAG in \citet{srinivasan2020fine} means multimodal ASR with global visual features. Multimodal (emb) in \citet{oneațua2022improving} fuses the two modality along the sequence dimension. ViLaS in \citet{han2023vilas} uses additional ASR datasets for pre-training and finetunes with mixed datasets.
    \label{comparison results}
\end{table}

\begin{table*}[htbp] 
    \renewcommand{\arraystretch}{1.1} % 调整行距
    \small %调整字体大小
	% \resizebox{\linewidth}{!}{
		\begin{tabular}{cccccccc}	 
			\hline
			\multirow{2}{*}{Dataset} & \multirow{2}{*}{Mask Ratio} & \multicolumn{2}{c}{Baseline} & \multicolumn{4}{c}{VHASR} \\
            \cmidrule(lr){3-4}\cmidrule(lr){5-8}
			&& $\mathrm{WER}$ ($\downarrow$)& $\mathrm{RR}$ ($\uparrow$)& $\mathrm{WER_{ASR}}$ ($\downarrow$)&$\mathrm{RR_{ASR}}$($\uparrow$)& $\mathrm{WER_{M_2}}$ ($\downarrow$)&$\mathrm{RR_{M_2}}$($\uparrow$) \\
			\hline
			\multirow{3}{*}{Flickr8k}&30\%&29.36&80.75 &27.39&83.22&\textbf{22.36}&\textbf{83.29}\\
            &50\%&46.79&69.80 &45.01&72.84&\textbf{38.35}&\textbf{73.38}\\
			&70\%&62.66&58.83 &63.43&60.60&\textbf{55.04}&\textbf{61.34}\\
            \hline
            \multirow{3}{*}{ADE20k}&30\%&24.79&92.02 &24.40&92.51&\textbf{19.96}&\textbf{92.60}\\
            &50\%&34.16&89.18 &32.95&89.86&\textbf{26.91}&\textbf{90.06}\\
			&70\%&42.30&86.33 &40.70&87.45&\textbf{33.39}&\textbf{87.46}\\
            \hline
            \multirow{3}{*}{COCO}&30\%&25.60&92.02 &24.23&92.85&\textbf{20.13}&\textbf{92.87}\\
            &50\%&35.59&89.42 &33.22&\textbf{91.05}&\textbf{27.06}&\textbf{91.05}\\
			&70\%&44.00&87.76 &41.35&89.26&\textbf{33.84}&\textbf{89.32}\\
            \hline	
   %          \multirow{3}{*}{OpenImages}&30\%&20.95&\textbf{96.45} &21.74&96.33&\textbf{17.50}&96.35\\
   %          &50\%&29.37&\textbf{94.58} &30.87&94.31&\textbf{24.49}&94.32\\
			% &70\%&37.08&\textbf{92.55} &38.94&92.27&\textbf{30.99}&92.28\\
   %          \hline	
		\end{tabular}
	% }
	\centering
    \caption{Experimental results of audio corruption with AWGN.}
    \label{audio corruption}
\end{table*}

The comparison results are shown in the Tabel \ref{comparison results}. Without vision information, Vilas \citep{han2023vilas} performs better than our VHASR since they have done sufficient pretraining. With vision information, VHASR's ASR performance has been significantly enhanced and it achieves the lowest WER. Obviously, our experimental results indicate that the incorporation of visual information aids in rectifying tokens for ASR transcription errors and decreasing WER. However, \citet{srinivasan2020fine}, \citet{oneațua2022improving} and \citet{han2023vilas} argue that the speech in Flickr8k is sufficiently clear, making it challenging to enhance transcription performance by incorporating additional information from other modalities. 
% This finding aligns with \citet{sun2016look}, \citet{srinivasan2020looking} and \citet{srinivasan2020multimodal}.

MAG \citep{srinivasan2020fine} utilize global visual features, which may introduce a significant amount of information unrelated to audio and potentially impact the model's ASR performance. They considered this issue and proposed MAOP, which utilizes multiple fine-grained image features extracted from object proposals. But in terms of clean Flickr8k, MAOP's performance is not as good as MAG's. \citet{oneațua2022improving} take a sequence of image features vectors from the layer preceding the global average pooling layer in the vision encoder, for leveraging more fine-grained characteristics of the image. However, they did not consider that some image vectors in the sequence have low correlation with the audio. Introducing these vectors fully into the backbone will still impact the model's recognition ability. \citet{han2023vilas} use ViT as a vision encoder and utilizes the image tokens for visual representation, which aligns with our approach. However, they do not reduce the weight of visual tokens with low importance, as we do. This resulted in the introduction of visual information not improving the recognition performance of the model. Compared to these works that use ordinary multimodal fusion approach, our proposed method, which injects visual modality information by vision hotwords, have made improvements in refining image representation and eliminating irrelevant image information. Therefore, our proposed model can enhance performance using visual features even when the dataset is of high quality and the baseline is strong.

\subsection{Audio Corruption}
\label{corruption}

To further demonstrate that introducing image information related to audio can reduce transcription errors in proposed model, we conduct an audio corruption experiment proposed by \citet{srinivasan2020looking}. We first use the timestamp prediction model proposed by \citet{Shi2023AchievingTP} to align audio and transcribed text. Then, we mask the words in the audio to a certain proportion by replacing the audio segments corresponding to the masked words with Additive White Gaussian Noise (AWGN). We use the recovery rate (RR) defined in \citet{srinivasan2020looking} to calculate the proportion of masked words recovered in the model transcription results. Unlike \citet{srinivasan2020looking}, our approach only masks the test data, while the training data remains unchanged.

We conduct this experiment on Flickr8k, ADE20k, and COCO, and the experimental results are shown in Table \ref{audio corruption}. In terms of baseline and ASR stream, regardless of the mask ratio, the ASR stream has lower WER and higher RR on all three datasets. This suggests that the jointly trained ASR stream exhibits stronger noise resistance and audio content prediction abilities compared to unimodal ASR. In terms of ASR stream and $M_2$, by incorporating image information, $M_2$ significantly reduces WER and enhances RR, as evidenced by the mask ratio across the three datasets. This indicates that image information can assist the model in capturing image-related words in audio, enabling the model to accurately transcribe these words even if their corresponding audio is masked. Furthermore, we can argue that on normal unmasked data, image information can assist the model in correcting words related to image but with transcription errors.

\subsection{Ablation Result}

To demonstrate that the refined image representation extracted by the method proposed in Section \ref{fine-grained} is more effective than the full image representation, we conduct the ablation experiments. The experimental results are presented in Table \ref{ablation results}. On four datasets, whether it is $M_1$ or $M_2$, the model using refined image representation has better performance. This not only shows the effectiveness of the method described in Section \ref{fine-grained} but also offers one of reasons why our model is stronger than other benchmarks.

\begin{table}[htbp] 
    \renewcommand{\arraystretch}{1.2} % 调整行距
    \small %调整字体大小
	\resizebox{\linewidth}{!}{
		\begin{tabular}{cccccc}	 
			\hline
			\multicolumn{2}{l}{\multirow{2}{*}{Dataset}} &\multicolumn{2}{c}{$\mathrm{WER_{M_1}}$ ($\downarrow$)} &\multicolumn{2}{c}{$\mathrm{WER_{M_2}}$ ($\downarrow$)} \\
			\cmidrule(lr){3-4}\cmidrule(lr){5-6}
			&&w/o refine& w refine& w/o refine & w refine \\
			\hline
			
			\multicolumn{2}{l}{Flickr8k} &3.88&\textbf{3.82}&3.67&\textbf{3.62}\\
			\multicolumn{2}{l}{ADE20k} &10.67&\textbf{10.38}&10.17&\textbf{9.80} \\
			\multicolumn{2}{l}{COCO} &10.46&\textbf{10.28}&9.64&\textbf{9.63} \\
            \multicolumn{2}{l}{OpenImages} &8.73&\textbf{8.58}&7.81&\textbf{7.73} \\
			\hline
						
		\end{tabular}
	}
	\centering
    \caption{ Experimental results of ablation studies.}
    % 'w refine' refers to VHASR using refined image representation, and 'w/o refine' refers to VHASR using full image representation. 
    \label{ablation results}
\end{table}

In order to showcase the strength of our baseline, we evaluate its ASR performance against Whisper. The experimental results are presented in the Table \ref{asrresults}. As the table shows, Whisper excels on F8k. This is attributed to: (1) Whisper's utilization of a large amount of data for pretraining, which we did not employ. (2) F8k being a high-quality dataset where many IBSR works achieve superior results without using visual information (refer to Table \ref{comparison results}). Nevertheless, our approach can enhance the ASR capability of the model by effectively leveraging visual information. In ADE20k, a dataset with more noise, our baseline demonstrates stronger noise resistance and performs better than Whisper. In essence, our baseline is on par with Whisper. Furthermore, our system's ASR module is adaptable and we will explore which ASR module can achieve optimal performance for VHASR in the future.

\begin{table}[htbp] 
	\small
    \renewcommand{\arraystretch}{1.2}    
	% \resizebox{\columnwidth}{!}{
		\begin{tabular}{ccccc} 
			\hline
			Model&Params&Trained&Flickr8k&ADE20k \\
            % \cmidrule(lr){2-2}\cmidrule(lr){3-3}
            % &CER($\downarrow$)&WER($\downarrow$)\\
			\hline
            Whisper&244M&\ding{51}&3.38&14.28\\ 
			Whisper&1.5B&\ding{53}&\textbf{3.05}&14.08 \\
            Baseline&220M&\ding{51}&3.86&10.51 \\
			\hline
            VHASR &333M&\ding{51}&\textbf{3.21}&\textbf{9.53} \\
            \hline
		\end{tabular}
	% }
    \centering
    \caption{WER of Whisper, our baseline and VHASR on FLickr8k and ADE20k. The 1.5B Whisper is version V3.}
    \label{asrresults}
\end{table}

\section{Conclusion}
We propose VHASR, a multimodal speech recognition system that utilizes vision hotwords to strengthen the model's speech recognition ability. Our system features a dual-stream architecture, consisting of an ASR stream and a VH stream that firstly transcribe separately and then combine their outputs. By leveraging vision hotwords, the VH stream concentrates on key visual information, allowing for precise transcription of words associated with images. In the merging phase, the VH stream assists the ASR stream in correcting any mis-transcribed words related to images, thereby ensuring high accuracy in the final transcription. We conduct comprehensive experiments on Flickr8k, ADE20k, COCO, and OpenImages, which showcase the effectiveness of vision hotwords and the robust ASR performance of VHASR.

% In this paper, we propose VHASR, a multimodal speech recognition system that utilizes vision hotwords to strengthen the model's speech recognition capability. To improve the effectiveness of cross-modal fusion, it calculates the similarity between different modalities. Through various experiments, we demonstrate that VHASR has powerful speech recognition performance.

\section*{Limitations}
The Limitations of VHASR include: (1) currently, VHASR can only introduce image information to enhance the model's speech recognition ability, which does not have sufficient versatility. In the future, we will enable VHASR to support input of audio-related text information (such as hotwords, titles) and video information, enabling the model to extract feature beneficial for speech recognition from multiple modal information, and building a more versatile multimodal speech recognition model. (2) we have only demonstrated that vision hotwords is a effective way to utilize image information, and there may be other applicable methods. We will design more in-depth experiments in the following work to explore more feasible ideas.

% in this paper, we has not yet deeply analyzed the reason of why image information can enhance model’s transcription ability, but only demonstrated at a coarse-grained level that introducing image information related to speech is beneficial. 

% \section*{Acknowledgements}
% This work was supported by the National Natural Science Foundation of China (No. 62306216, No. 72074171, No. 72374161), the Natural Science Foundation of Hubei Province of China (No. 2023AFB816), the Fundamental Research Funds for the Central Universities (No. 2042023kf0133). 

\bibliography{custom}

\clearpage
\appendix
\section{Appendix}
\subsection{Case Study}
In Section \ref{corruption}, we demonstrated that VHASR can use image information to correct words which is related to images and has transcription errors. In this section, we will use examples to explain how VHASR achieves this.

\begin{figure*}[htbp]
	\centering 
	\includegraphics[scale=0.8]{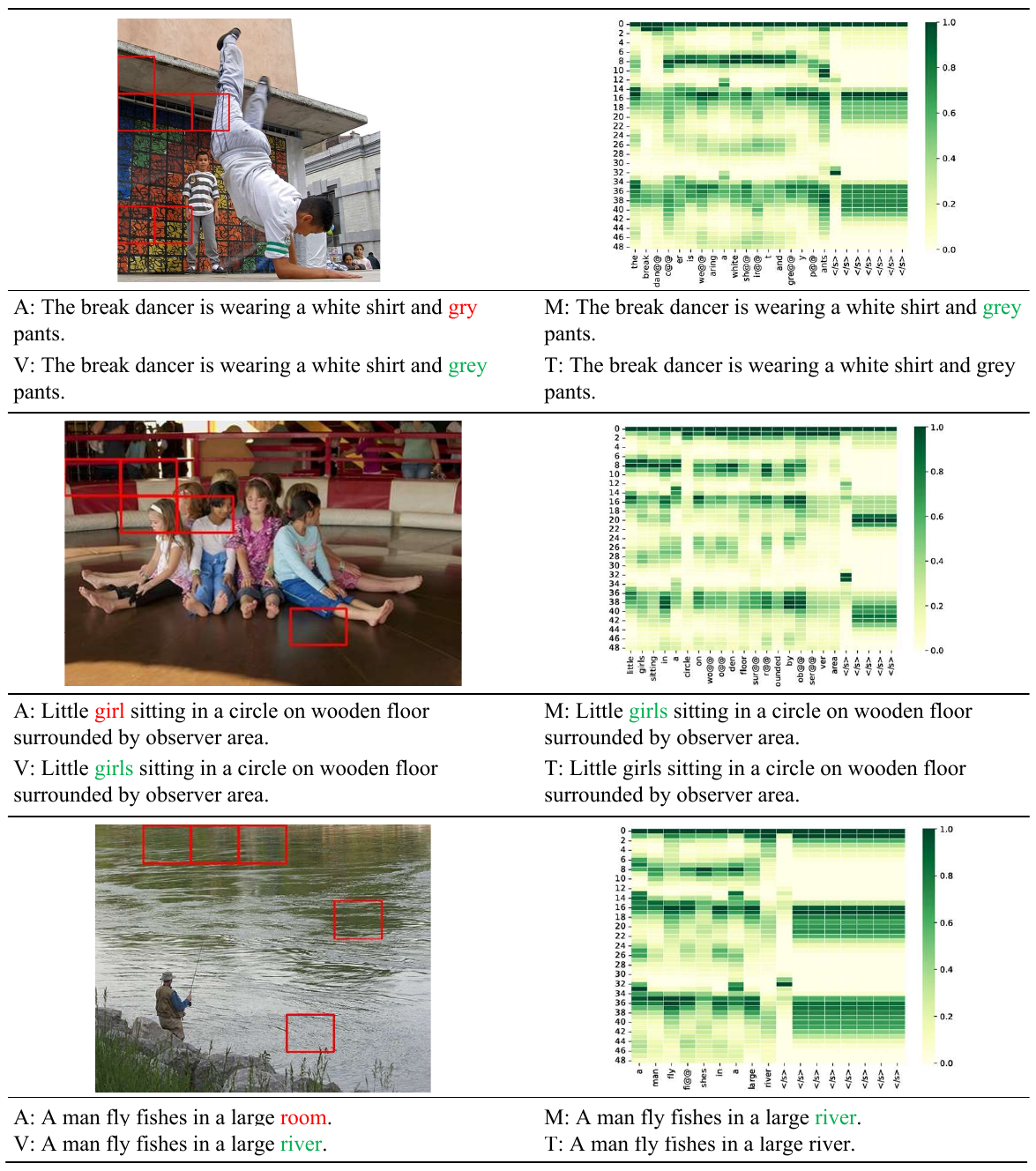} 
	\caption{Three examples about how VH stream helps to rectify ASR stream's error.} 
    \label{case}
\end{figure*}

Figure \ref{case} shows three examples from Flickr8k. "\emph{A}" refers to the transcription of the ASR stream, "\emph{V}" refers to the transcription of the VH stream, "\emph{M}" refers to the transcription obtained by $M_3$, and "\emph{T}" refers to the real transcription. We extract the attention score matrix from the last layer of the VH decoder and create a heatmap. The horizontal axis of the heatmap represents the subtoken, while the vertical axis represents the number of vision hotwords. We identify the subtokens that are transcribed incorrectly by the ASR stream but corrected by the VH stream. Then, we extract the top 5 vision hotwords that have the highest attention scores with them. Chosen vision hotwords are marked on the original image. 

\begin{figure*}[htbp]
	\centering 
	\includegraphics[scale=0.8]{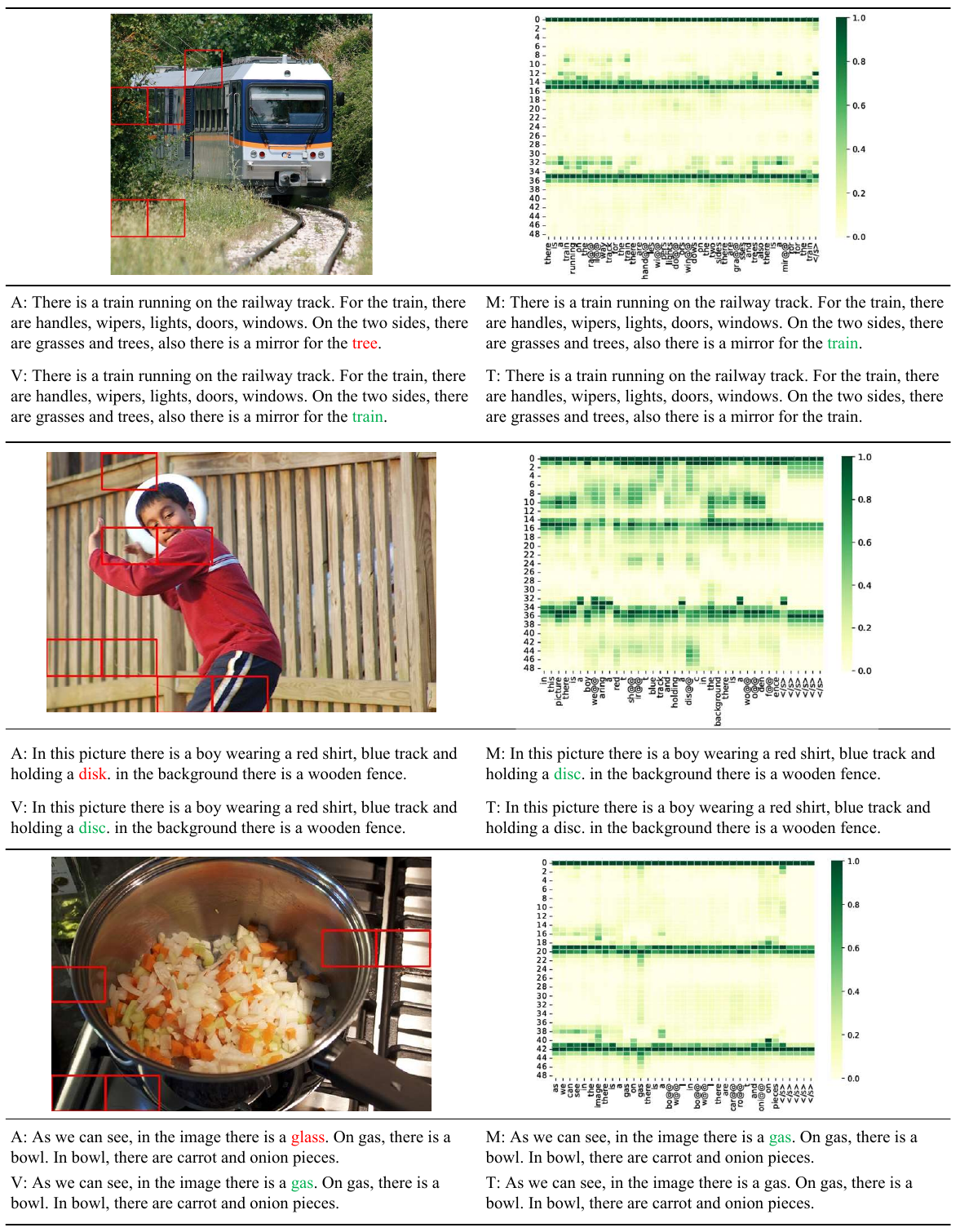} 
	\caption{More examples about case study.} 
    \label{case2}
\end{figure*}

In the first example, the ASR stream incorrectly transcribes "\emph{grey}" as "\emph{gry}", while the VH stream doesn't make this mistake. The combination of the two streams helps correct the error. specifically, the subtokens corresponding to "\emph{grey}" focus on six vision hotwords, five of which are background, and one includes the grey pants of the dancer. Therefore, the vision encoder successfully extracts information about "\emph{grey}" and helps the VH stream transcribe "\emph{grey}" accurately. Furthermore, by merging the ASR stream and VH stream with $M_3$, error in the ASR stream is rectified. In the second example, the ASR stream incorrectly transcribes "\emph{girls}" as "\emph{girl}", which was also corrected by the accurate VH stream. Among the vision hotwords corresponding to "\emph{girls}", three are related to background, and two include the heads of the girls, so the VH stream successfully identified "\emph{girls}". In the third example, the ASR stream incorrectly transcribes "\emph{river}" as "\emph{room}", but the VH stream correctly transcribes "\emph{river}" by utilizing the information about "\emph{river}" contained in the vision hotwords. By merging, the VH stream helps correct error in the ASR stream. These examples are not unique, and the same phenomenon occurs in many utterances. In Figure \ref{case2}, we show another three examples from COCO for readers' reference.

Although the VH stream of VHASR has less speech recognition ability than the ASR stream, it can extract features from key vision hotwords and capture keywords in transcription, thereby correctly identifying words that may be difficult for the ASR stream to recognize. After token-by-token merging based on visual-token similarity, the VH stream can correct some transcription errors in the ASR stream, leading to a more accurate transcription.

\end{document}